\documentclass[a4paper,11pt]{article}
\pdfoutput=1

\usepackage{jcappub}
\usepackage[table,usenames,dvipsnames]{xcolor}
\usepackage{ifthen}
\usepackage{amsmath}
\usepackage{verbatim}
\usepackage{graphicx}
\usepackage{natbib}
\usepackage[table]{xcolor}

\notoc

\begin{document}

\title{Searching for concentric low variance circles in the cosmic microwave
 background}
\author[a,b,1]{Adam DeAbreu,\note{Corresponding author.}}
\author[b]{Dagoberto Contreras,}
\author[b]{and Douglas Scott}

\affiliation[a]{Department of Physics, Simon Fraser University,\\ Burnaby,
 BC, V5A 1S6 Canada}
\affiliation[b]{Department of Physics \& Astronomy, University of British
 Columbia,\\ Vancouver, BC, V6T 1Z1 Canada}

\emailAdd{adeabreu@sfu.ca}
\emailAdd{dagocont@phas.ubc.ca}
\emailAdd{dscott@phas.ubc.ca}

\date{\today}

\abstract
{In a recent paper, Gurzadyan \& Penrose claim to have found directions in
the sky around which there are multiple concentric sets of annuli with
anomalously low variance in the cosmic microwave background (CMB).
These features are presented as evidence for a particular theory of the
pre-Big Bang Universe. We are able to reproduce the analysis these authors
presented for data from the {\it WMAP\/}
satellite and we confirm the existence of these apparently special directions
in the newer {\it Planck\/} data. However, we also find that these features
are present at the same level of abundance in simulated Gaussian CMB skies,
i.e.,\ they are entirely
consistent with the predictions of the standard cosmological model.}

\maketitle
\flushbottom

\section{Introduction}
\label{sec:intro}

The cosmic microwave background (CMB) can be used to constrain the physics of
the Universe on the largest scales and at the earliest times. Many searches
have been undertaken for anomalous signatures on the CMB sky
(see ref.~\cite{PlanckIS} and references therein);
however, such searches are usually
phenomenological in nature, because there are rarely definitive predictions
from theoretical models. One exception to this is the assertion that in a
particular cyclic cosmology picture proposed by Roger Penrose \cite{CCC},
one expects to see rings of low variance on the CMB sky~\cite{GP10}.
Indeed Gurzadyan \& Penrose claimed to
have found such a signature in maps of the CMB made by the Wilkinson Microwave
Anisotropy Probe ({\it WMAP}, \cite{Jarosik}). However, this claim was
quickly countered by three papers (Moss et al.~\cite{Moss2011}, Wehus et
al.~\cite{Wehus2010}, and Hajian~\cite{Hajian2011}) pointing out that such low
variance rings occur frequently in simulated CMB skies. As stressed by
Moss et al.\ in particular, the fact that the CMB anisotropy power spectrum
is known to have structure on the scale of the examined rings gives greater
dispersion among their properties; this assures the
existence of some low variance rings, even for a temperature field with
entirely uncorrelated phases.

In a more recent paper (ref.~\cite{GP13}, hereafter GP13)
Gurzadyan \& Penrose assert that the conformal cyclic cosmology model should
give {\it multiple\/} concentric rings of low variance around points on the
sky. In {\it WMAP\/} data, they claim to find an abundance
of sets of three or more concentric rings, and use this to suggest that there
may be evidence on the CMB sky for previous ``aeons'' in the history of the
Universe, before the Big Bang. This is such an extraordinary claim that it
certainly merits further investigation.  As stressed by Penrose\footnote{Talk available at:
\url{http://physics.princeton.edu/cmb50/home.shtml}\,.} \cite{CMB50},
the newer multiple-rings prediction was not explicitly checked in the
previously published studies. We note that Wehus et
al.~\cite{Wehus2010}, did look for several simple forms of concentric low
variance circles in the CMB, however, they did not look for the specific
prediction presented in GP13 (not to the fault of the authors of \cite{Wehus2010}
since GP13 was released after their analysis). In fact the test proposed in GP13 is
specific enough that it is relatively straightforward to apply it to both real
and (Gaussian random) simulated CMB skies, in order to determine whether our
actual sky possesses an anomalously high abundance of concentric rings.

GP13 analysed maps from the {\it WMAP\/} 7-year data release. In particular
they searched for directions around which there are annuli with low variance in
the temperature field on the sky, finding instances of multiple concentric
annuli with low variance. For each direction considered they calculated the
standard deviation, $\sigma$, for 28 annuli of width $0.5^\circ$, with radius
in the range $2.5^\circ$--$16^\circ$.  An annulus around a given direction
was considered to have low variance if its value of $\sigma$ was at least
$15\,\mu$K below the average among all annuli centred on the same direction.
As an additional criterion,
if two or more exactly adjacent annuli for the same direction
had low variance they were instead treated as a single, wider, low
variance annulus.  Amongst all directions, those having three or more low
variance annuli surrounding them were considered to be special directions. A
simple $20^\circ$ cut about the Galactic plane ($|b| < 20^\circ$) was made and
no directions were selected within this region, nor were any pixels within the
masked region considered for the purpose of calculating the annuli variances.
Following this approach,
the abundance of these low variance directions was stated to be more than a
statistical feature of the CMB sky, but rather claimed as cosmological in
nature and possible evidence for pre-Big Bang phenomena.  In addition,
the substantial decrease in the number of low variance annuli at large radius
was asserted to be a further signature of previous aeons.

In this paper we have tried to reproduce the analysis of GP13, also applying
it to improved data available from the {\it Planck\/} satellite, as well as
to simulations.  In doing so we attempted to assess whether the abundance of
sets of concentric rings is unexpectedly high.  In a later section we examine
different approaches to determining the threshold for selecting an annulus as
having low variance, and in doing so we address the issue of how the
abundance is expected to depend on the radius of the annulus.

\section{Finding low variance rings}
\label{sec:rings}

\begin{figure}[tbp]
\centering
\begin{minipage}{0.49\textwidth}
\includegraphics[width=\columnwidth,angle=0]{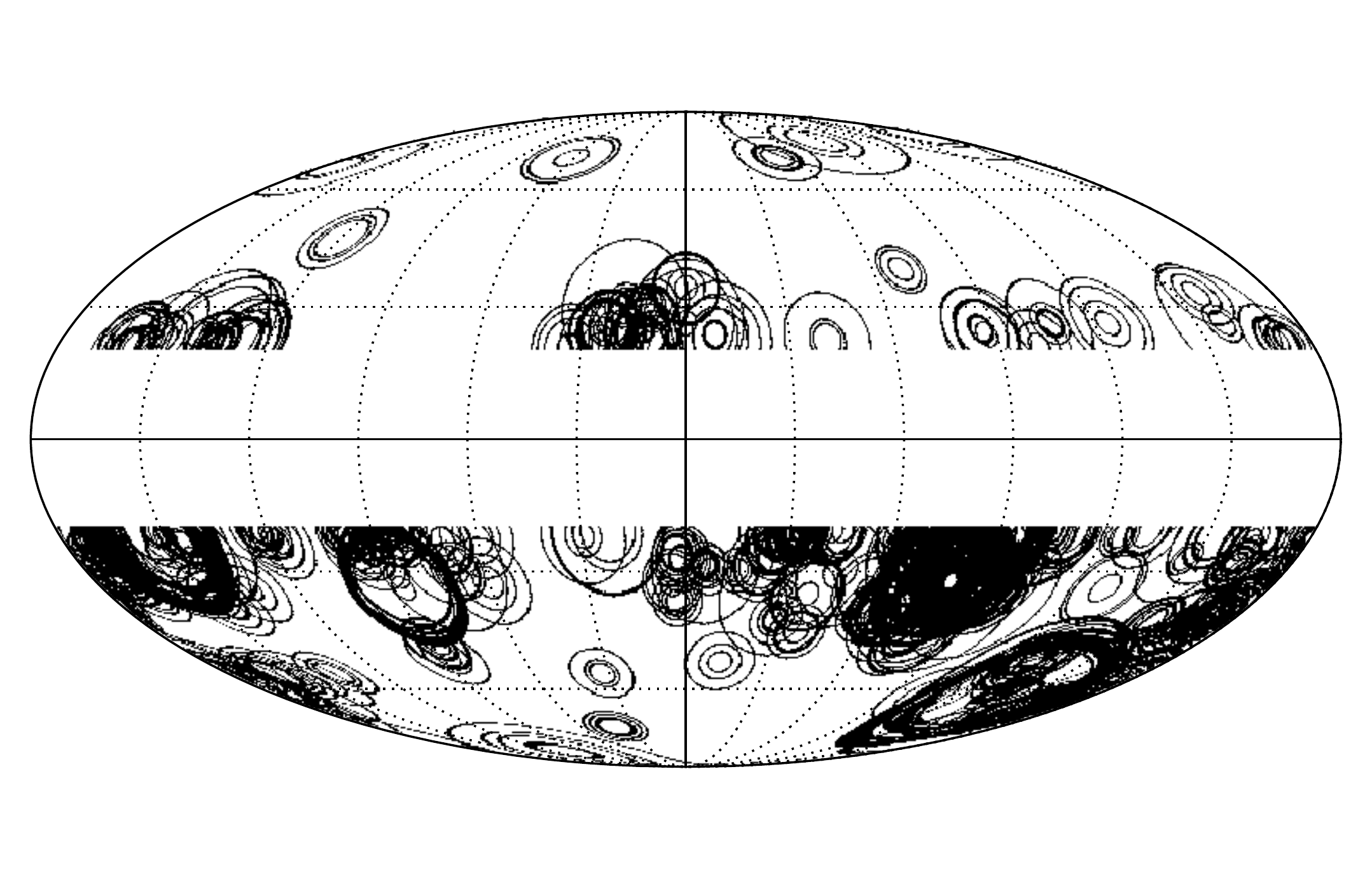}
\end{minipage}
\begin{minipage}{0.49\textwidth}
\includegraphics[width=\columnwidth,angle=0]{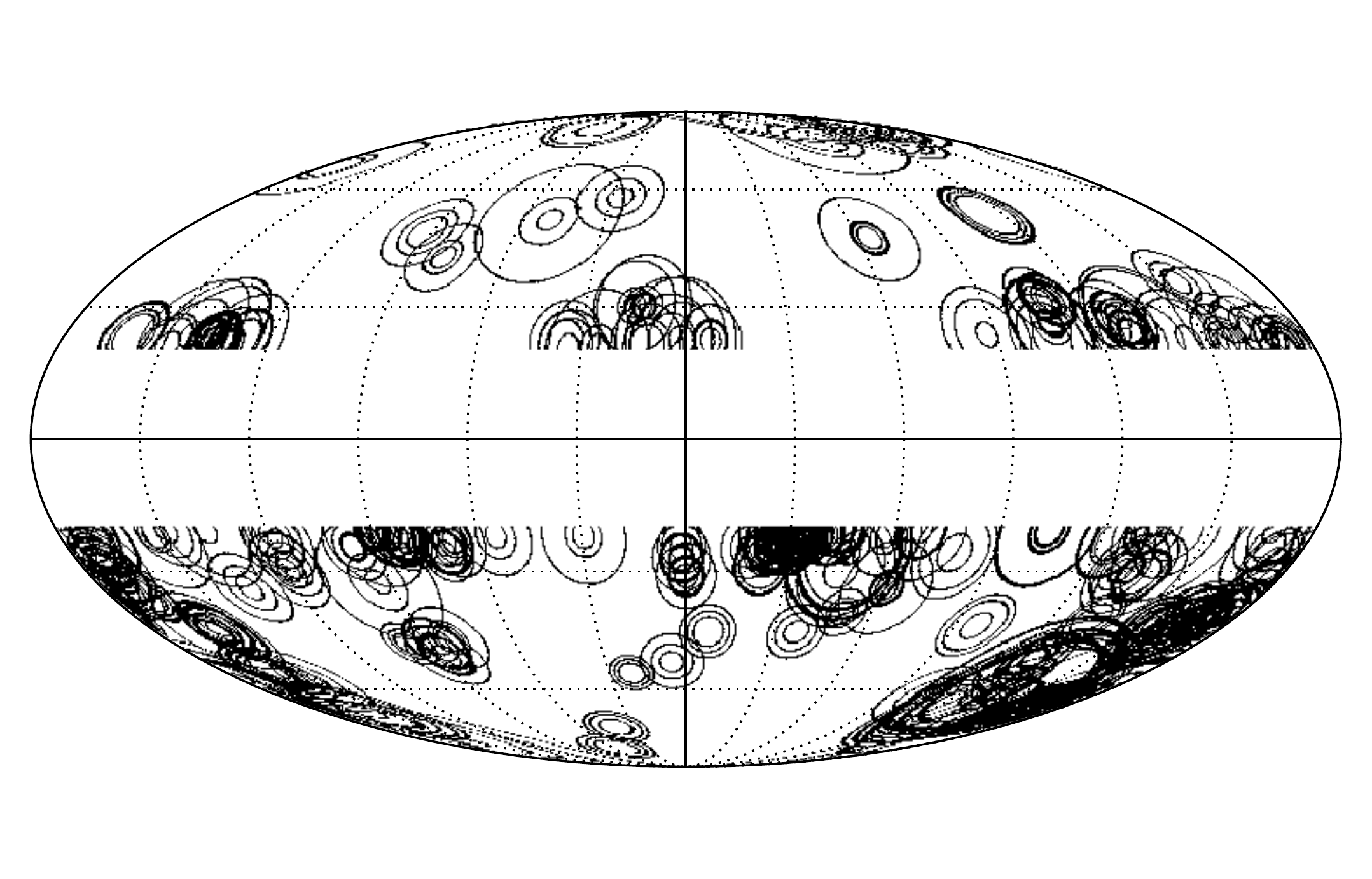}
\end{minipage}
\caption{\label{fig:wmap_dirs} (left) All-sky map showing the
distribution of low variance directions (i.e.,\
directions with three or more low variance annuli around them, using the same
criteria described in GP13) in the
{\it WMAP\/} W-band data set, smoothed with a 20 arcminute FWHM Gaussian.
There are 228 directions satisfying the search criteria. (right) The equivalent
map using the {\tt Commander} data set smoothed with a 20 arcminute FWHM Gaussian.
There are 166 directions satisfying the search criteria. Note that the differences between the
data sets are due to the non-negligible, inhomogeneous, noise present in the {\it WMAP\/}
data.}
\end{figure}

We first tried to reproduce the result presented in GP13, using the same method
and the same {\it WMAP\/} data set.  The {\it WMAP\/} W-band data
are used,\footnote{Obtained from the LAMBDA site:
\url{http://lambda.gsfc.nasa.gov/}\,.}
smoothed with a 20 arcminute full-width at half-maximum (FWHM) Gaussian beam.
For the set of directions considered we used the centres of $N_{\rm side}=32$
pixels (12{,}288 directions) in the {\tt HEALPix}~\cite{Gorski}
pixelization scheme, skipping those that were within $20^\circ$ of the
Galactic plane.
The annuli are bound by $R$ and $R+\Delta R$, with $\Delta R = 0.5^\circ$ and
$R$ in the range $2.5^\circ$--$16^\circ$ (this range and annulus width
were chosen to match GP13). Figure~\ref{fig:wmap_dirs} (left) shows the distribution
over the sky of low variance annuli for directions with three or more such
annuli.  We found 228 low variance directions (defined as a direction with three or more annuli with
$\sigma$ being $15\,\mu$K or more below the average for that direction),
distributed in a manner
that closely matches the distribution seen in figure~3b of GP13.
We note that the visually striking asymmetry in the distribution
of low variance circles in the {\it WMAP\/} data is not present to the same degree
in the {\it Planck\/} {\tt Commander} data.
Despite the appearance of more rings in one hemisphere than the other, we find
that about $25\,\%$ of simulations produce a distribution of low variance directions with a higher dipole
asymmetry than seen in the data. We conclude that the asymmetry is not statistically
significant.
The difference between the {\it WMAP\/} and {\tt Commander} data is most
likely due to the inhomogeneous noise that is non-negligible at these scales.

Next we investigated the same statistic in the newer data from the
{\it Planck\/} satellite~\cite{PlanckOverview}, specifically using the
{\tt Commander} component-separated map~\cite{PlanckComponent}.
The results of this paper are not sensitive to the choice of component-separation algorithm.
The {\tt Commander} data set was selected, since it is the preferred map for large and intermediate angular scales~\cite{PlanckComponent},
which are the most relevant in this analysis. Note that although
{\it Planck\/} is more sensitive and has higher resolution, at the angular
scales relevant for this study, the {\it Planck\/} and {\it WMAP\/} data were
expected to be very similar. To speed up the computations this data set was
downgraded to $N_{\rm side}=512$ and smoothed with a 20 arcminute FWHM
Gaussian beam and the same analysis as performed on the {\it WMAP\/} data
set was repeated. We applied the same procedure to simulations created from the
{\it Planck\/} best-fit $\Lambda$CDM angular power spectra
\cite{PlanckLikelihood} generated at $N_{\rm side}=2048$, downgraded
to $N_{\rm side}=512$ and smoothed with a 20 arcminute FWHM Gaussian.

\begin{figure}[tbp]
\centering
\includegraphics[width=\columnwidth,angle=0]{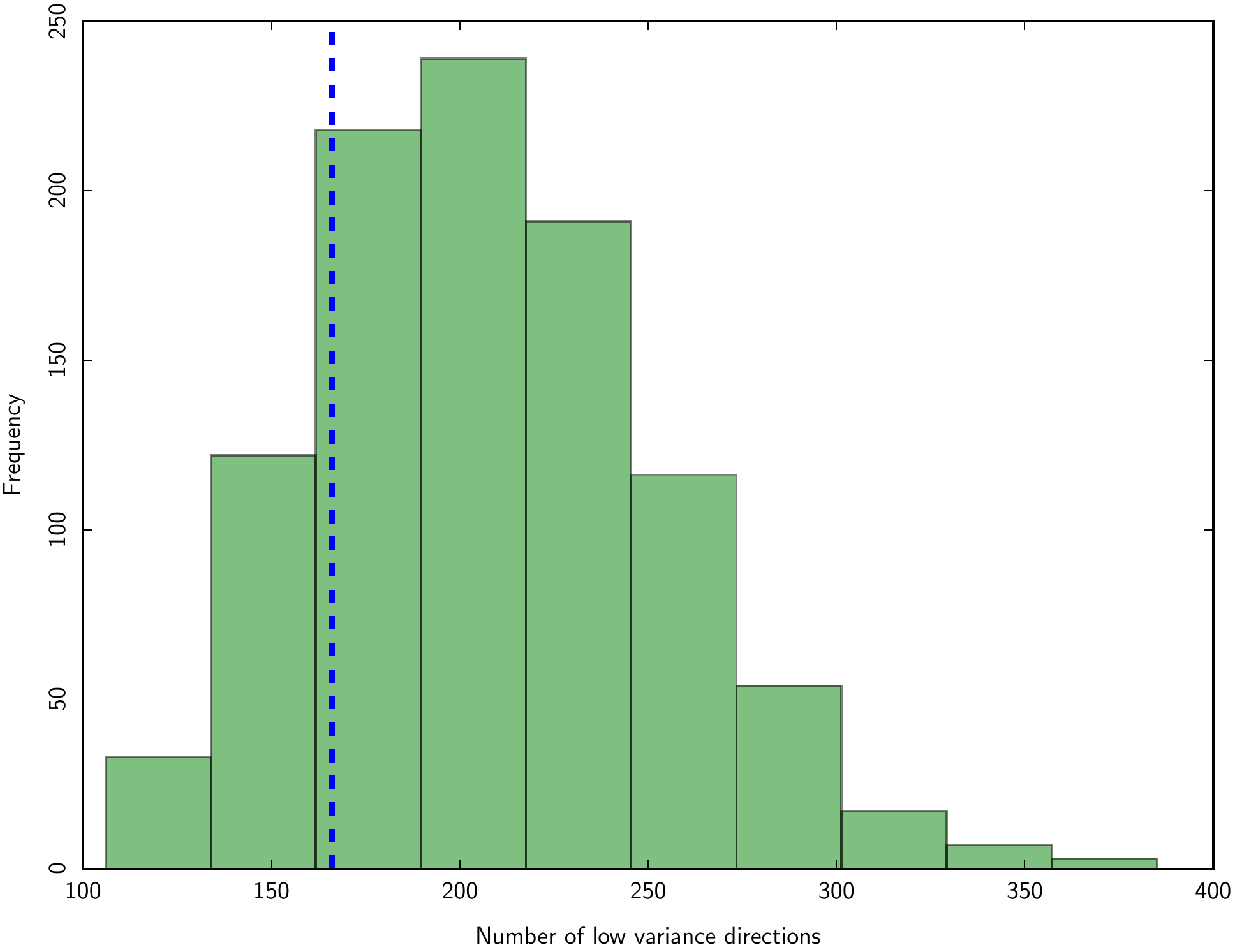}
\caption{\label{fig:linemean_hist} Number of low variance directions (i.e.,\
directions with three or more low variance annuli around them) in the
{\it Planck\/} data (blue vertical line) compared to 1000
simulations (histogram). The threshold used here for a low
variance annulus about a given direction is that the standard deviation
should be at least $15\,\mu$K below the
average value across all annuli for that direction (as defined in GP13).
The data do
not prefer high numbers of low variance directions.}
\end{figure}

Figure~\ref{fig:wmap_dirs} (right) shows the distribution
over the sky of low variance annuli for directions with three or more such
annuli.  We found 166 low variance directions (defined as a direction with three or more annuli with
$\sigma$ being $15\,\mu$K or more below the average for that direction),
distributed in a manner
that closely matches the distribution seen in the {\it WMAP\/} data (figure~\ref{fig:wmap_dirs}, left).
Figure~\ref{fig:linemean_hist} shows a histogram of the number of directions
with three or more concentric low variance rings found in simulations.  It is
clear that we see just as many directions in simulations as in the real sky; in
other words the presence of these concentric low variance annuli is not
significant when compared to simulations.

\section{Defining low variance}
\label{sec:defining}

\begin{figure}[tbp]
\centering
\includegraphics[width=\columnwidth,angle=0]{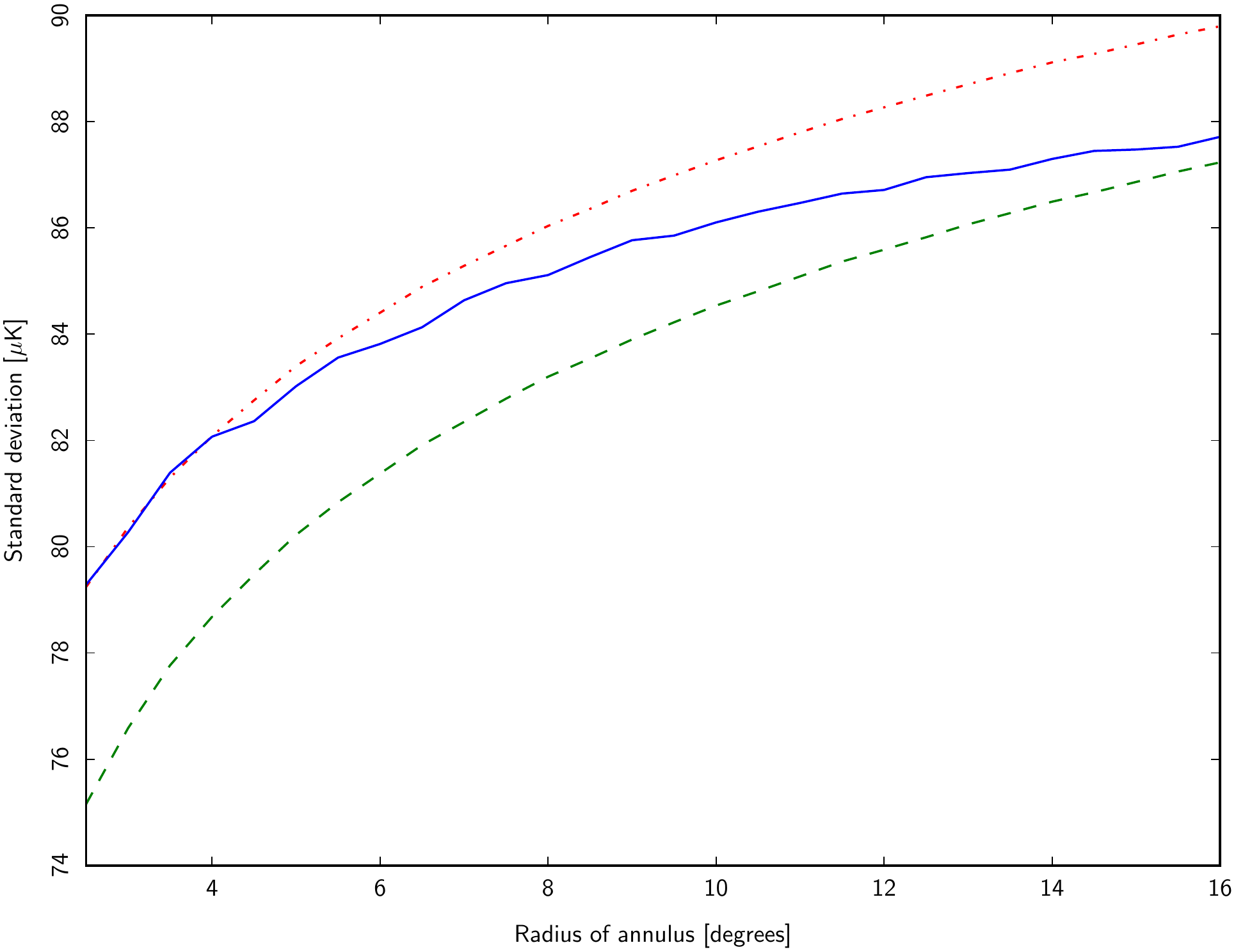}
\caption{\label{fig:radial_std} Standard deviation of annuli bounded by $R$ and
$R+\Delta R$, with $\Delta R = 0.5^\circ$, averaged across 12{,}288 directions
for {\it Planck\/} data (solid blue), ignoring points within a $\pm20^\circ$
Galactic latitude cut.  We also plot the average of simulations without
masking (dot-dashed red) and
for simulations with each ring half-masked by a horizontal cut (dashed green).
This shows that the expected variance depends both on the amount of masking
and on the radius of the annulus.  Hence it is important to track these effects
in the variance estimation. }
\end{figure}

We next turned our attention to the particular definition of a ``low variance
annulus'' used by GP13, namely a single average variance value per direction in
order to define a threshold.
The average radial profile of the annular standard deviation of simulations
(created as mentioned before) and {\it Planck\/} data are shown in
figure~\ref{fig:radial_std}.  The variance increases as the annular radius
increases, as described in Moss et al.\ \cite{Moss2011},
this being the result of the scale-dependence of CMB
anisotropies.  This fact makes a simple single value average a poor choice when
defining a threshold for low variance.  An additional complication is
that the variance of an annulus changes depending on what fraction of the
annulus is masked. This is because the masking yields a horizontal cut to the
annulus, excluding some fraction of the circle and thus effectively cutting out
some large-scale modes that contribute to the variance.
Figure~\ref{fig:radial_std} shows the average radial profile for unmasked
annuli and partially masked annuli.  It is clear that a simple single-value
average will not capture the tendency for masked rings to have a lower
variance.

To address this we used simulations to devise an adaptive threshold based on
the direction and annular radius considered. For 1000 simulations, masked in
the same manner as the data (a $\pm20^\circ$ Galactic cut) the variance was
calculated for each annulus bounded by $R$ and $R+\Delta R$, with
$\Delta R = 0.5^\circ$ and $R$ in the range $2.5^\circ$--$16^\circ$, around
12{,}288 directions defined by the centres of $N_{\rm side}=32$ pixels,
ignoring points within the masked region.  By averaging across all simulations
we obtained an estimate of the expected variance for every annulus around every
direction considered.  Since the simulations were masked in the same way as
the data, these estimates provided a variance value to compare to for any given
direction and annular radius encountered in the analysis, which accounts for
the masking due to the Galactic cut.

We then repeated the search for low variance annuli, using the criterion that
the standard deviation should be at least $15\,\mu$K below this comparison
value for each direction.  We stuck to the same choice of threshold,
simply because it was used in GP13; we make no claims to the optimality of
this selection.

Before showing the results for this ``adaptive threshold'' we will discuss
another choice, which is based on the spread of the variance
for each direction and annular radius.  Smaller annuli are composed of a
smaller number of pixels than larger annuli and thus will have a larger spread
in variance. In the same manner that we used simulations to create a comparison
variance value for each direction and annular radius, we can also calculate a
variance of this annular variance, hereafter $\sigma_\text{v}$. Thus, rather
than using $15\,\mu$K as the low variance criterion, we can use a cut
of $2\sigma_\text{v}$ below the comparison
variance for each direction and annular variance.  The choice of
$2\sigma_\text{v}$ was made since it effectively lies close to $15\,\mu$K,
therefore keeping quantitatively in line with the threshold used by GP13.
Again, we do not claim that this is the best choice, but it certainly
accounts for the nature of the annuli better than a flat $15\,\mu$K cut.
The specific choice of this threshold will alter the number of low variance
directions found in the data; however, since the same analysis is performed on
the simulations the assessment of significance is {\it not\/} sensitive to this
choice.

\begin{figure}
\centering
\includegraphics[width=\columnwidth,angle=0]{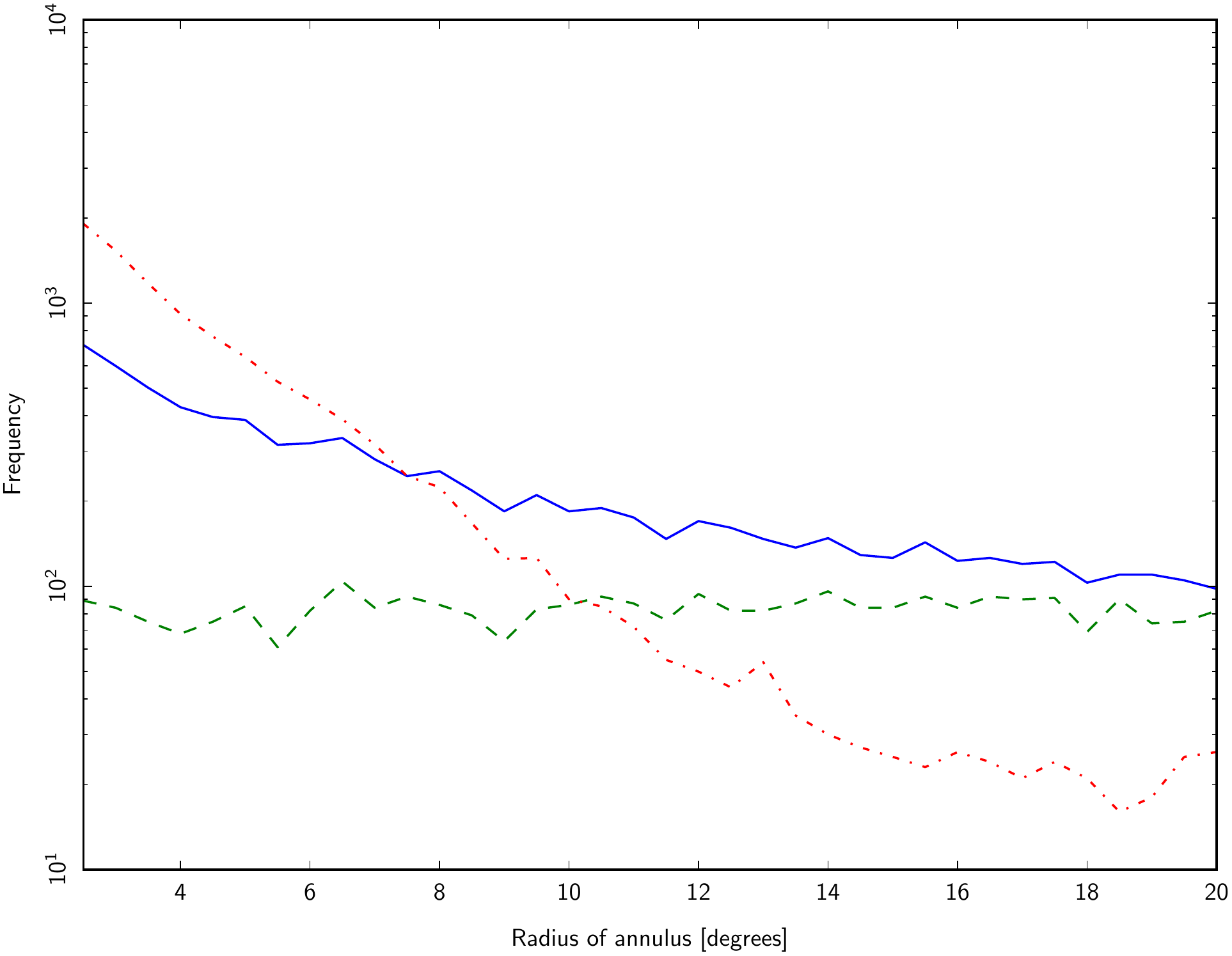}
\caption{\label{fig:low_dist} Frequency with which annuli bounded
by $R$ and $R+\Delta R$, with $\Delta R = 0.5^\circ$, are classified as ``low
variance'' within the {\it Planck\/} data for the three
threshold choices defined previously: a single average value for each
direction, as used by GP13 (dot-dashed red); the adaptive comparison curve
defined using simulations, with a $15\,\mu$K threshold (solid blue), and with a
$2\sigma_\text{v}$ threshold (dashed green). Comparing the dot-dashed red
curve to the others shows that the lack of low variance rings found at large
radii is a consequence of the choice in threshold.}
\end{figure}

The effect of picking different ways to calculate the threshold can be seen in
figure~\ref{fig:low_dist}.  GP13 claim that the lack of low variance annuli at
large radii is of crucial importance for their particular model.  However, this
appears simply to be due to their particular definition of the threshold -
because the annular variance increases monotonically with radius it becomes
less likely that a large radius annulus will be classified as having low
variance by their selection procedure (see the red curve in
figure~\ref{fig:low_dist}).  The other two choices of threshold shown in the
figure do not show this large drop in the number of low variance annuli at
large radius, since they compensate for the fact that the annular variance
increases with radius.  In particular, the combination of the adaptive
comparison and the $2\sigma_\text{v}$ threshold shows no preference for any
particular radius (see the green curve in figure~\ref{fig:low_dist}).  The
equivalent curves in figure~\ref{fig:low_dist} for simulations show the same
behaviour.

\begin{figure}
\centering
\includegraphics[width=\columnwidth,angle=0]{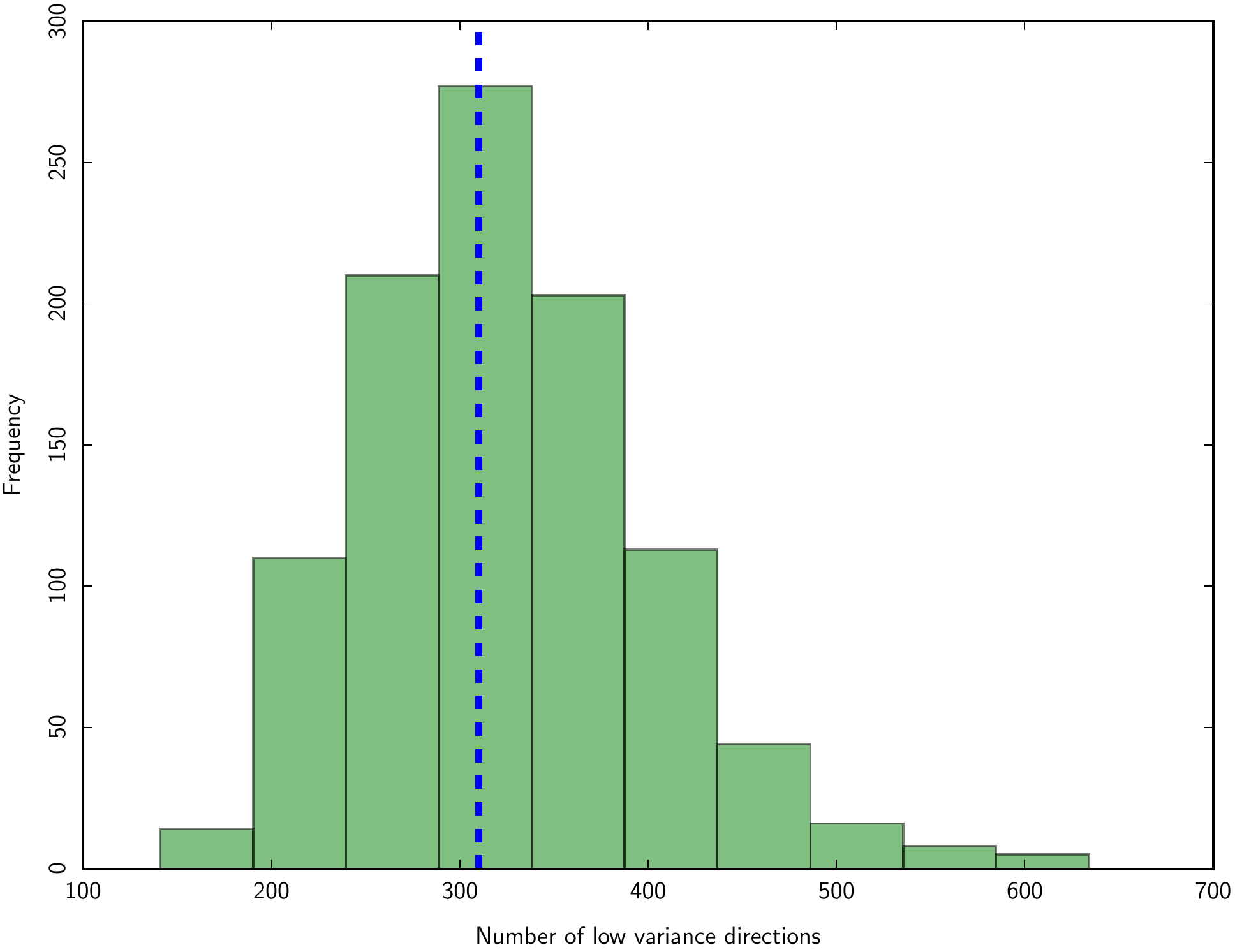}
\caption{\label{fig:reg_sim_hist}  Number of low variance directions
(directions with three or more low variance annuli around them) in the
{\it Planck\/} data (blue vertical line) compared to 1000
simulations (histogram). The threshold here is a ring with a standard
deviation at least $15\,\mu$K below a comparison value generated from
simulations based on each annular radius and fraction masked.
The data do
not prefer high numbers of low variance directions.}
\end{figure}

We would suggest that these other choices of threshold are better suited to
defining low variance rings.  But regardless of the threshold used, the
remainder of the analysis remains the same as previously described. Figure~\ref{fig:reg_sim_hist} shows
the number of directions with three or more concentric low variance rings found
in data and simulations using the adaptive threshold with a $15\,\mu$K cut for
a low variance annulus.  Once again, the number of low variance directions seen
in simulations matches that seen in the real sky.
In addition to the $2\sigma_\text{v}$ threshold, we also looked across a
range of possible $\sigma_\text{v}$ levels; the results remain the same for all values considered, i.e.,
the data
do not prefer high numbers of low variance directions compared to simulations.

GP13 further claim that there is a crucial dependence on circularity, seen in a lack of low variance ellipses
compared to low variance circles. They carry out this analysis using a procedure that rotates the sky to produce approximate ellipses from circles. However, this will entirely change the temperature correlations on the sky, i.e., the expected variance properties, and in particular the variance profile as a function
of radius (see figure~\ref{fig:radial_std}).
We explicitly tested this and found that the variance profile does indeed depend on the amount
of ``twist'' applied in the procedure.
Thus, this is not a fair comparison, particularly when using the threshold defined by GP13.
To properly study ellipses on the sky would require calculating the expected variance as a function of
the full parameter space of
ellipse properties and then searching over those properties in the data. Such a task would be much more computationally intensive than the analysis that we have carried out here.

\section{Conclusions}
\label{sec:conclusions}

In summary we have confirmed that there are many sets of concentric
low-variance rings in the CMB sky; however, the number of such directions
(regardless of search criteria used) containing such rings are consistent with
what one would expect from Gaussian random skies containing the usual CMB
anisotropy power spectrum.  As shown in figure~\ref{fig:reg_sim_hist}, in a
perfectly standard $\Lambda$CDM universe, about half of the observers in
separate Hubble patches will see a greater abundance of concentric low-variance
rings in their CMB sky than we see in ours. Furthermore, we have shown that the
apparent drop in low variance rings at large annular radius is merely a result
of the particular search criterion used and has no statistical significance.

Therefore we
conclude that based on searching for concentric low-variance rings in our CMB
sky we have found \emph{no\/} evidence for previous cycles in the history of
our Universe prior to the Big Bang, as might be predicted by the conformal
cyclic cosmology model.

The analysis presented here adds another test to an ever growing list that
the standard $\Lambda$CDM model passes with ease.  We stress that
when performing such tests care must be taken to assess the significance of
the results using simulations, since we must always ask
how likely it is that a random realization of a $\Lambda$CDM sky will exhibit
a signature with similar characteristics to the one we see in the real sky.
Despite the null result reported here, it is nevertheless important to
continue to carry out further searches for anomalies in the large-scale
pattern of CMB anisotropies, since it is still a promising
avenue for exploring physics beyond the standard cosmology.

\acknowledgments
This research was supported by the Natural Sciences and Engineering Research
Council of Canada.  We thank Jim Zibin for his assistance and for many helpful
discussions.

\bibliographystyle{JHEP.bst}
\bibliography{bullseye}

\end{document}